\documentclass[useAMS,usenatbib]{mn2e}

\usepackage{amssymb}
\usepackage{multirow}
\usepackage{graphicx}
\usepackage{aas_macros}

\def\msun{$M_{\odot}$}

\def\ergsec{\hbox{erg s$^{-1}$}} 
\def\ergcm{\hbox{erg s$^{-1}$ cm$^{-2}$}}

\def\Chandra{\emph{Chandra}}

\title[M28 Neutron Star Atmosphere Composition]{Neutron star atmosphere composition: the quiescent, {low-mass} X-ray binary in the globular cluster M28}
\author[M. Servillat et al.]{M. Servillat$^{1,2}$\thanks{E-mail: mservillat@cfa.harvard.edu}, C. O. Heinke$^{3}$, W. C. G. Ho$^{4}$, J. E. Grindlay$^{1}$, J. Hong$^{1}$, \newauthor{M. van den Berg$^{5,1}$, and S. Bogdanov$^{6}$}\\
$^{1}$Harvard-Smithsonian Center for Astrophysics, 60 Garden Street, Cambridge,~MA~02138, USA\\
$^{2}$Laboratoire AIM (CEA/DSM/IRFU/SAp, CNRS, Universit\'e Paris Diderot), CEA Saclay, Bat. 709, 91191 Gif-sur-Yvette, France\\
$^{3}$Department of Physics, University of Alberta, CCIS 4-183, Edmonton AB, T6G 2E1, Canada; Ingenuity New Faculty\\
$^{4}$School of Mathematics, University of Southampton, Southampton, SO17 1BJ, United Kingdom\\
$^{5}$Astronomical Institute Anton Pannekoek, University of Amsterdam, Science Park 904, 1098 XH Amsterdam, the Netherlands\\
$^{6}$Columbia Astrophysics Laboratory, Columbia University, 550 West 120th Street, New York, NY 10027-6601, USA
}

\begin{document}

\date{Accepted .... Received ...; in original form ...}

\pagerange{\pageref{firstpage}--\pageref{lastpage}} \pubyear{...}

\maketitle

\label{firstpage}

\begin{abstract}
Using deep \Chandra\ observations of the globular cluster M28, we study the quiescent X-ray emission of a neutron star in a low-mass X-ray binary in order to constrain the chemical composition of the neutron star atmosphere and the equation of state of dense matter.
%The X-ray emission does not show significant variability between or during each observation over six years.
We fit the spectrum with different neutron star atmosphere models composed of hydrogen, helium or carbon. 
The parameter values obtained with the carbon model are unphysical and such a model can be ruled out. 
Hydrogen and helium models give realistic parameter values for a neutron star, and the derived mass and radius are clearly distinct depending on the composition of the atmosphere.
The hydrogen model gives masses/radii consistent with the canonical values of 1.4~\msun\ and 10~km, and would allow for the presence of exotic matter inside neutron stars. On the other hand, the helium model provides solutions with higher masses/radii, consistent with the stiffest equations of state.
%Unless the composition of the accreting material can be further constrained, measurements from quiescent neutron stars in globular cluster that showed low radii or masses might instead harbor an helium atmosphere and have a larger radius or mass.
%Unless one can constrain the type of donor, i.e. white dwarf or non-evolved star, we show that the constraints
Measurements of neutron star masses/radii by spectral fitting should consider the possibility of heavier element atmospheres, which produce larger masses/radii for the same data, unless the composition of the accretor is known independently.
\end{abstract}


%--------------------
\section{Introduction}

Neutron stars (NS) are composed of the densest form of matter known to exist in our Universe, providing us with a unique laboratory to study cold matter at supra-nuclear density. 
%, and thus provide a unique laboratory for exploring the properties of cold matter at supranuclear density. 
%Different equations of state ( EOSs) of dense matter predict different maximum masses and different mass-radius relationships.
%what is the symmetry energy and thus the proton fraction in the core, the behavior of superfluidity among neutrons and protons, or the conductivity of the NS crust. 
For example, it is still not well understood whether exotic condensates occur in the NS core. The chemical composition of the outer envelope is also uncertain, as well as the symmetry energy, the behavior of superfluidity among neutrons and protons, and the conductivity of the NS crust. 
Measuring the masses or radii of these objects can lead to useful constraints on the dense matter equation of state (EOS), and give insights of the composition of NSs (see \citealt{Lattimer:2010p8404} for a recent review). 
% \citealt{Lattimer:2001p6716,Lattimer:2007p6670}

%Methods to measure mass and radius, qLMXB...

%\cite{Steiner:2010p7695}
Masses have been determined very accurately for a few dozen NSs in binaries containing pulsars (for a recent compilation, see \citealt{Lattimer:2010p6671}). Measured masses for a variety of NS systems range from 1.0 to 2.5~\msun, with a canonical mass of 1.4~\msun. In particular, the high NS mass measured with high precision for the pulsar PSR J1614-2230 ($1.97\pm0.04$~\msun) brings doubts on the presence of hyperons or free quarks in NS interiors \citep{Demorest:2010p6698}, though those solutions are not completely ruled out \citep[e.g.][]{Weissenborn:2011p8410,Massot:2012p7975}.

%However, no radius information is available for these systems.
For systems in which the NS is hot enough to emit detectable X-rays from the surface, due to youth or accretion, the X-ray spectrum can be used to constrain both the radius and mass of the NS \citep{Lattimer:2010p8404,Ozel:2010p6772,Steiner:2010p7695,Galloway:2012p8635}.
%accretes matter from a nearby companion, nuclear processes in the crust and envelope of the NS provide additional observables that can be used to constrain its mass, but also its radius \citep[e.g.][]{Ozel:2010p6772}.
The mass and radius of isolated NS or ones in transient low-mass X-ray binaries (LMXBs) can be inferred from spectral modeling if their distances are accurately determined. In the case of accreting NS transients located in globular clusters (GCs), relatively accurate distances are known (errors of $\sim$5\%, see e.g. \citealt{Krauss:2003p7606}). Even though, the uncertainties for individual NS measurements are still large \citep{Rutledge:2002p6656,Gendre:2003p1003,Heinke:2006p1069,Webb:2007p649,Guillot:2011p6829}, but the ensemble of observations can ultimately improve constraints on the dense matter EOS \citep[e.g.][]{Steiner:2010p7695}.

%Mass measurements of high accuracy for a number of radio pulsars in close binary systems (often with inferred NS companions) are consistent with a range of NS masses between 1.25 and 1.45 Ms ( Thorsett & Chakrabarty 1999). Fundamental constraints on the NS interior structure can be achieved by measurement of the gravitational redshift from the NS surface (Cottam et al. 2002). Finally, it should be possible to derive constraints on the radius of NSs from spectral fits to their X-ray emission if the temperature, composition of atmosphere, and distance to a NS are known and the magnetic field is sufficiently weak so as not to affect the opacity or temperature distribution on the NS surface. These requirements can be fulfilled for X-ray observations of quiescent low-mass X-ray binaries (qLMXBs) containing NSs, particularly those lo- cated in GCs to which the distance is well known ( Brown et al. 1998; Rutledge et al. 2002b).

%These NS systems generally show soft spectra, consisting of a thermal, blackbody-like component, and possibly a harder component extending to higher energies, usually fitted with a power law of photon index 1--2 (although some systems are dominated by the harder component; Campana et al. 2002; Wijnands et al. 2005a). The thermal component, if fitted by a blackbody, produces inferred radii too small for theoretical NS size estimates. 

It has been shown that the surface of a weakly magnetic ($B < 10^{10}$~G) NS should be chemically very pure and dominated by the lightest element present as the heavier elements settle out of the atmosphere within seconds to minutes \citep{Alcock:1980p7979,Brown:2002p8015}.
%\citep{Romani:1987p7705}. 
%Therefore, if the accretion rate falls below $10^{-13}$~\msun~yr$^{-1}$ (Brown et al. 1998), 
If there is accretion after the NS formation, the atmosphere could be composed of hydrogen --H-- or helium --He-- as heavier elements are expected to be destroyed via nuclear spallation reactions \citep{Bildsten:1992p7901,Chang:2004p8411}. A fraction of the incident He also suffers spallation reactions and may reform through fusion reactions \citep{Bildsten:1993p7907}. The ratio of H to He is thus not well determined.
If no accretion takes place or if all lighter elements are burned, heavy elements are expected (\citealt{Chang:2010p7934}, and references therein).
%thermonuclear reactions occur after accretion, heavy elements are expected.
%For accreting quiescent LMXBs (qLMXBs), depending on the composition of the accreted material, the atmosphere could be mostly composed of hydrogen, helium or heavier material (e.g. if the companion is a main sequence star or a white dwarf).
%For low-metallicity atmospheres, the emitted spectrum is harder than the corresponding blackbody spectrum. The spectra emitted from high-metallicity atmospheres (carbon, oxygen or iron) are closer to a blackbody distribution, but show strong absorption structures in the energy range observable with X-ray telescopes \citep{Romani:1987p7705,Ho:2009p7676}.

Different NS atmosphere models have been developed, but most recent work for low magnetic fields has focused on a pure H model, such as the ones developed by \citet{Zavlin:1996p6827}, \citet{Gansicke:2002p7900}, or \cite{Heinke:2006p1069}. 
%The latter covers a wide range of surface gravities in addition to
The latter model, NSATMOS, was further developed to represent atmospheres of pure He, carbon, nitrogen, oxygen or iron \citep{Ho:2009p7676}. In particular, such models were used for the low magnetic field NS located at the center of the Cassiopeia A supernova remnant, which was shown to harbor a carbon atmosphere \citep{Ho:2009p7676}.
% This suggests that there is nuclear burning in the surface layers and also identifies the compact source as a very young (330-year-old) NS.
%Moreover, such a system in a GC

%At densities above a few times the equilibrium density of nuclear matter, models predict the existence of exotic components such as pion or kaon condensates or unconfined quarks (e.g., Lattimer & Prakash 2007). The exciting idea that NSs may con- tain exotic forms of matter makes them of prime interest, not only for astrophysics but for physics in general.

%Quiescent LMXBs found in globular clusters (GCs) are routinely used to produce precise measurements of NS radii. The distance to GCs are generally known with precisions of the order of 5\%--10\% and the overabundances of NS binary systems in GCs (Hut et al. 1992) make these objects ideal targets to pursue in searches of qLMXBs capable of providing useful constraints on the dense matter EoS.

%\cite{Chaboyer:2008p7667}
%the review by \cite{Krauss:2003p7606} reviews 8 diﬀerent methods to determine the distances to GCs, and concludes that these various methods imply an uncertainty in the distance modulus to GCs of $\pm$0.12 mag, or $\pm$6\% in the distance
%\cite{Testa:2001p7589} The HB luminosity is estimated at VHB 15.55 +/- 0.10, in agreement with Davidge et al. 1996 (VHB 15.5)

%In this letter, we test the composition of the atmosphere of a NS in the GC M28.
The GC M28 (NGC 6626) is located at a distance of $D=5.5\pm0.3$~kpc (from \citealt{Harris:1996p7568,Harris:2010p7572}, using measurements in \citealt{Testa:2001p7589}) at ${\mathrm{RA} = 18^{h}24^{m}32.81^{s}}$ and ${\mathrm{Dec} = -24^{\circ}52'11.2''}$ (J2000).
%close to the Galactic center ($l=7.80^{\circ}$, $b=5.58^{\circ}$, \citealt{Harris:1996p7568,Harris:2010p7572}). 
%M28 is a compact GC...
The reddening toward M28 is ${E(B-V) = 0.42\pm0.02}$ \citep{Testa:2001p7589}, implying a H column density of $N_{\rm H} = (2.33\pm0.12)\times10^{21}$~cm$^{−2}$ (using \citealt{Predehl:1995p5533} for conversion).

\citet{Becker:2003p734} have previously reported on a set of $\sim$40~ks \Chandra\ X-ray Observatory ACIS-S observations of M28 (ObsIds 2683, 2684, 2685).
% which probed down to a luminosity of $L_{X} \sim 6 \times 10^{30}$~\ergsec. 
They suggested that the luminous, soft \Chandra\ source numbered 26 in their work (IAU-approved source name CXOGlb J182432.8-245208) is a transiently accreting NS in a LMXB in quiescence (qLMXB). 
We keep the name source 26 throughout the text.
Fitting its spectrum with a H atmosphere model (NSA model, \citealt{Zavlin:1996p6827}) yielded the effective temperature $T_{\infty}\sim90$~eV and the projected radius $R_{\infty}\sim14.5$~km with the mass set to 1.4~\msun, corresponding to a NS temperature $T_{\rm NS}\sim125$~eV and radius $R_{\rm NS}\sim10.4$~km.

In this paper we focus on this NS qLMXB in M28 using one of the deepest \Chandra\ observations of a GC to date and recent atmosphere models with different chemical compositions. We present the data in Section~\ref{data}, and a variability and spectral analysis in Section~\ref{results}. We finally discuss the results and implications in Section~\ref{discuss}.

%--------------------
\section{Data}
\label{data}

\begin{figure*}
\centering
\includegraphics[width=\textwidth]{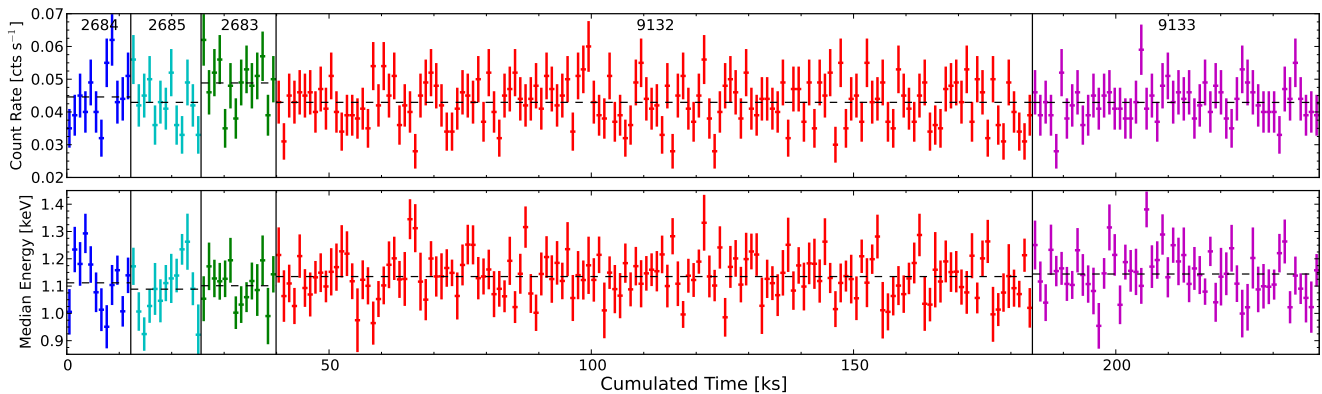}\\
\caption{X-ray lightcurve of the M28 \Chandra\ source 26. The five ObsIds are presented in consecutive pannels. The count rate and the median energy are calculated in the 0.3--6~keV energy band.}
\label{fig:lcs}
\end{figure*}

\begin{table*}
\caption{Properties of the X-ray observations of the M28 \Chandra\ source 26.\label{tab:prop}}
\begin{tabular}{cccccccccc}
\hline
%\tablehead{\colhead{ObsId} & \colhead{Date} & \colhead{Exp.Time} & \colhead{Counts} & \colhead{Rate} & \colhead{Median Energy} & \colhead{Mean Energy} & \colhead{$\chi^{2}$/dof} & \colhead{KS prob.} & \colhead{Frac. rms} \\ 
{ObsId} & {Date} & {Exp.Time} & {Counts} & {Rate} & {Median Energy} & {Mean Energy} & {$\chi^{2}$/dof} & {KS prob.} & {Frac. rms} \\ 
 & & [ks] & & [$10^{-2}$ c/s] & [keV] & [keV] & & \% & \% \\
\hline
2684 & 2002-07-04 &  12.2  &  561 & 4.4$\pm$0.2 & 1.11$\pm$0.03 & 1.2$\pm$0.5 &  10.7/13   & 22  & $<$23 \\
2685 & 2002-08-04 &  13.4  &  569 & 4.3$\pm$0.2 & 1.08$\pm$0.02 & 1.2$\pm$0.5 &   12.8/14   & 32  & $<$24 \\
2683 & 2002-09-09 &  14.3  &  684 & 4.9$\pm$0.2 & 1.10$\pm$0.03 & 1.2$\pm$0.5 &   12.8/14   & 48  & $<$21 \\
9132 & 2008-08-07 & 144.2 & 6182 & 4.29$\pm$0.05 & 1.13$\pm$0.01 & 1.2$\pm$0.5 & 120.6/144 & 52  & $<$12 \\
9133 & 2008-08-10 &  54.7  & 2336 & 4.3$\pm$0.1 & 1.14$\pm$0.01 & 1.2$\pm$0.5 &  26.0/56  & 45  & $<$21 \\
%[0.5ex] \tableline \\[-2ex]
\hline
All    &                  & 238.8 & 10332 & 4.29$\pm$0.04 & 1.13$\pm$0.01 & 1.2$\pm$0.5 & 173.3/200 & 38  & $<$11 \\
\hline
\end{tabular}
\end{table*}

The Chandra data set used here was already presented by \citet{Bogdanov:2011p3141}. Two long observations were acquired on 2008 August 7 (ObsId 9132) and 2008 August 10 (ObsId 9133) for 144 and 55 ks, respectively (PI: J. Grindlay). 
%VFAINT, TIMED, frame time = 3.141s
%CALDBVER= '3.4.5   '  
%CIAO 4.3
%For both exposures, the optical center of the cluster at RA = 18 24 32.89 and Dec = --24 52 11.4 (Shawl \& White 1986) was positioned 1′ off-axis from the nominal aim point of the back-illuminated ACIS-S3 CCD, configured in VFAINT telemetry mode. 
We also included the three observations presented by \citet[][ObsIds 2683, 2684, and 2685, PI: W. Becker]{Becker:2003p734}.
All observations were taken with ACIS-S in VFAINT telemetry mode, and TIMED read mode with a frame time of 3.1~s and reading time of 0.041~s.
The data re-processing, reduction, and analysis were performed in the same way as presented by \citet{Bogdanov:2011p3141}, but using the more recent {CIAO~4.4 and CALDB~4.4.8}.
In particular, we did not used the back-ground cleaning algorithm specific to the VFAINT telemetry mode as this procedure tends to reject real source counts for relatively bright sources.

%ACIS Extract to optimize the source extraction
We used the CIAO tool wavdetect and the script PWDetect \citep{Damiani:1997p7712} to detect source candidates in the field of view. Subsequently, we employed the IDL script ACIS Extract \citep{Broos:2010p7524} to confirm the validity of the source detections and refine the source positions. 
This lead to the detection of 101 sources inside the half mass radius of M28, and 35 inside the core radius, with a minimum of 4 counts per source.

The qLMXB candidate, source 26 of \citet{Becker:2003p734}, is detected with a total of 10332 counts ($\sim$0.043~cts~s$^{-1}$) in the 0.3--6~keV energy band. 
We extracted the events of the target in a 1.25\arcsec\ radius region, enclosing 95\% of the total source energy for an effective energy of 1~keV. 
We estimated that the contamination from close-by detected sources is $<$1\%.
The background was extracted from three 24\arcsec\ radius regions surrounding the GC core radius, making sure the regions are source free. 
{The data is clean from background flares, except in ObsId 2683 where the background flux reaches 5 times its median value for 4~ks over the 14~ks observation. However, this corresponds to a negligible contamination of $\sim$0.5 count in the aperture of the target (0.1\% of the target counts in that ObsId), we thus kept all the data in this obsid.}

We used \textit{dmextract} to extract a lightcurve with a 1000~s binning (minimum of 30 counts per bin), and \textit{dmcopy} to extract an event list for the source and background. Using \textit{specextract}, which includes an energy dependent aperture correction (\textit{arfcorr}), we extracted a spectrum and generated response files for each dataset. The spectra were binned with a minimum of 50 counts to use $\chi^2$ statistics for the fitting. We  discarded bins flagged as bad, and bins below 0.3~keV where the response of the instrument is not calibrated {(calibration down to 0.3~keV with ACIS-S3 has been clearly improved in CALDB 4.4\footnote{http://cxc.harvard.edu/cal/memos/contam\_memo.pdf}).}

%We extracted the emission from a region that encloses 90\% of the total source energy at 1.5 keV. To permit spectral fitting in XSPEC the extracted source counts in the 0.3--8 keV range were grouped in energy bins so as to ensure at least 15 counts per bin. The background was taken from three source-free regions in the image around the cluster core.

%--------------------
\section{Results}
\label{results}

\subsection{Light curve and search for variability}
\label{var}

We show in Figure~\ref{fig:lcs} the lightcurve of the target for all five observations with a 1000~s binning, as well as the median energy of the source in each bin. 
We report the lightcurve characteristics and variability tests in Table~\ref{tab:prop}.
For each lightcurve we computed the $\chi^{2}$ of a fit with a constant and performed a Kolmogorov-Smirnov (KS) test of variability. 
The source showed no significant variability in all Chandra observations. There is no feature in the light-curve that would indicate a periodic variability, or flares, and we found no correlation between the flux and median energy of the source as a function of time.
We also give the fractional rms (root mean square) of the light-curve. As the observed variability is consistent with noise, a 3$\sigma$ limit on the fractional rms is given in Table~\ref{tab:prop}.

The possible 4$\sigma$ increase of flux reported for ObsId 2683 compared to 2684 and 2685 \citep{Becker:2003p734} is not observed in this work. The mean count rate is slightly higher (about 2$\sigma$) in observation 2683, but the scatter around the mean for this ObsId is not unusual, as can be seen in Figure~\ref{fig:lcs}. Similarly, we found no significant change in the mean energy. There may be an increase of the median energy between 2002 and 2008 of $\sim$0.03~keV, which is marginal (less than 2$\sigma$).
%{With age and accumulation of radiation on the detector, the charge transfer inefficiency (CTI) is expected to rise, leading to a slight decrease of the energy scale, by $\sim$0.01~keV for the back-illuminated chip S3 over such a period\footnote{http://space.mit.edu/$\sim$cgrant/AAS11.poster.letter.pdf}, contrary to the possible increase observed. However, the lightcurve is consistent with the source being constant over that period.}
%while the energy scale of the ACIS-S detector is generally observed to decrease with age of $\sim$0.01~keV over such a period. However, the significance of the change is lower than 2$\sigma$, consistent with the source being constant over that period.

\subsection{Spectral fitting}
\label{fit}

%\subsubsection{Hydrogen atmosphere model}
\label{fitnsatmos}

\begin{figure}
\centering
\includegraphics[width=\columnwidth]{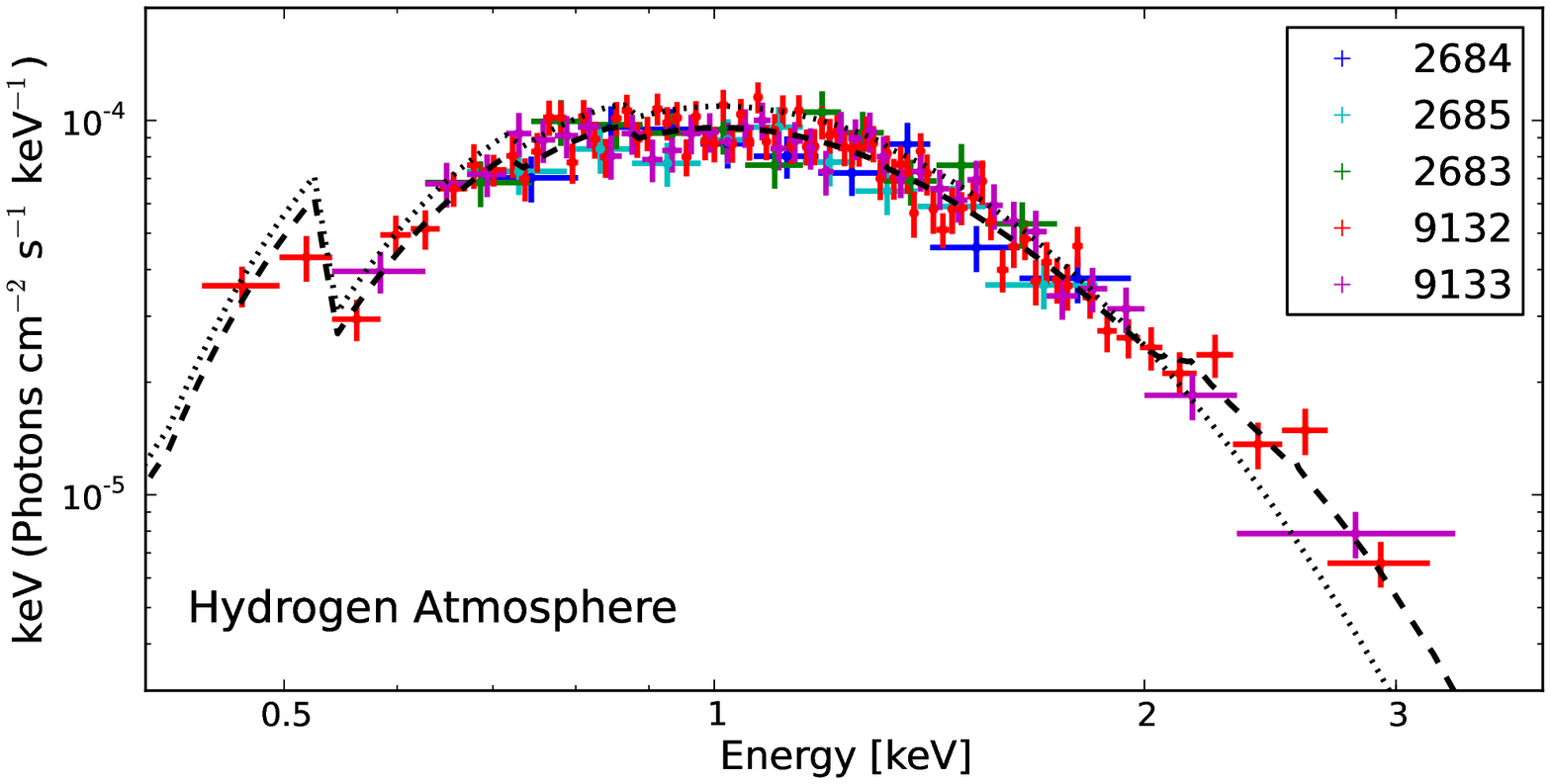}\\
\includegraphics[width=\columnwidth]{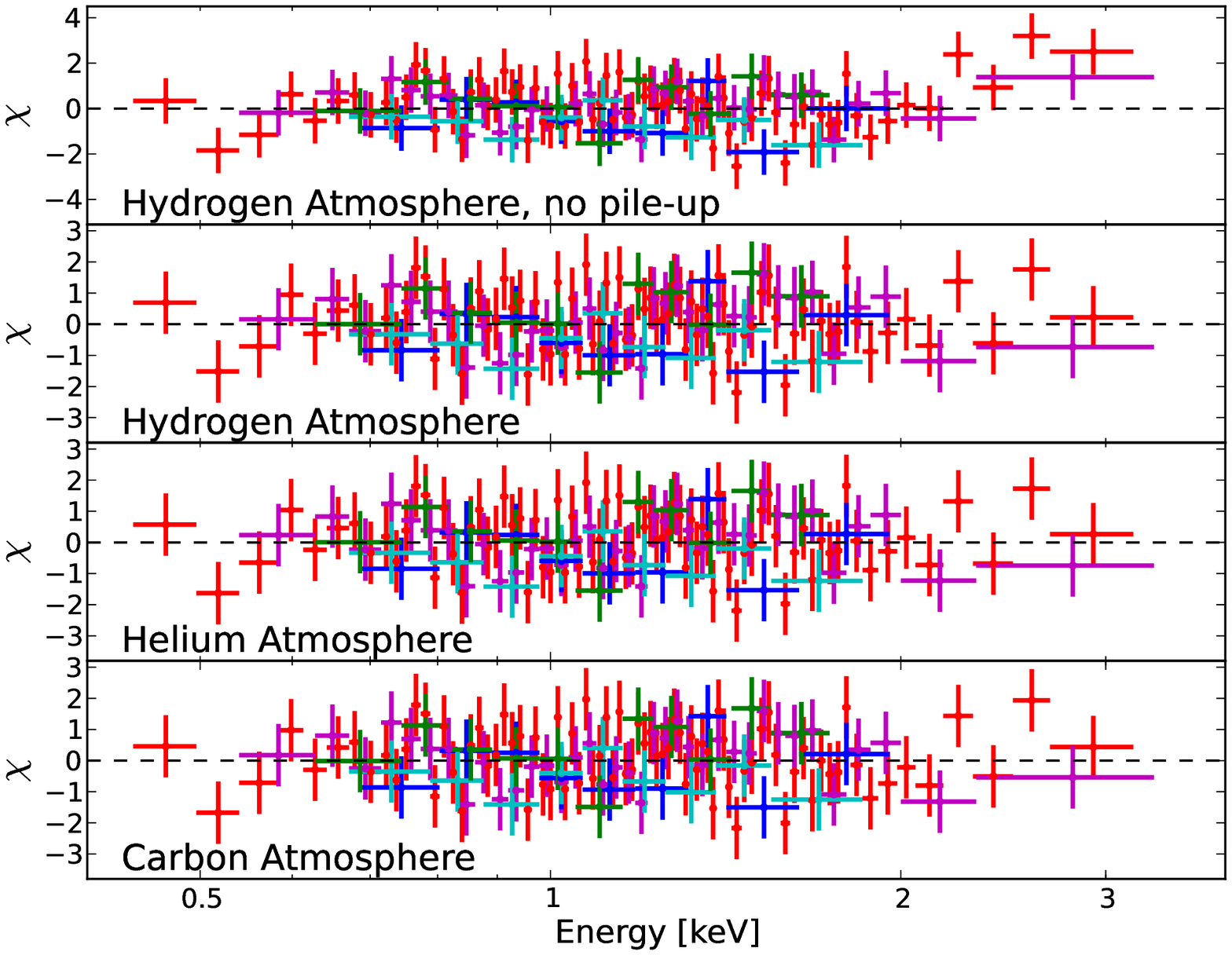}
%\caption{Spectra of the M28 NS \Chandra\ source 26 fitted with a H (top) and a He atmosphere (bottom). The best fit models are reported as dashed lines. The dotted lines shows the model without pile-up.}
\caption{Spectrum of the M28 \Chandra\ source 26 fitted with a NS atmosphere. The best fit model for the pure H atmosphere is reported as a dashed line and the dotted line shows the model without pile-up. Residuals are shown for the initial H atmosphere fit without pile-up and for different chemical compositions.}
\label{fig:sp}
\end{figure}

We fitted simultaneously the five spectra extracted from the five different epochs with Xspec 12.7.0e \citep{Arnaud:1996p4268}, using the pure H atmosphere model NSATMOS \citep{Heinke:2006p1069} and a photoelectric absorption $N_{\rm H}$ along the line of sight (TBABS, with abundances from \citealt{Wilms:2000p2492}). We first tied all parameters between the different datasets, which seems a reasonable assumption as the source did not show significant variability between the different epochs (section~\ref{var}). We fixed the distance to 5.5~kpc and the normalization to 1 (i.e. we assume that all the NS surface is emitting). We also kept at first the mass and radius to the canonical values, 1.4~\msun\ and 10~km respectively.
This returned a marginally acceptable fit with a reduced $\chi^{2}$ ($\chi^{2}_{\nu}$) of 1.27 and 144 degrees of freedom (dof). Data bins at energies higher than 2~keV show systematic positive residuals, mainly visible for the ObsId 9132 spectrum (see Figure~\ref{fig:sp}).

\begin{figure*}
\centering
\includegraphics[width=\columnwidth]{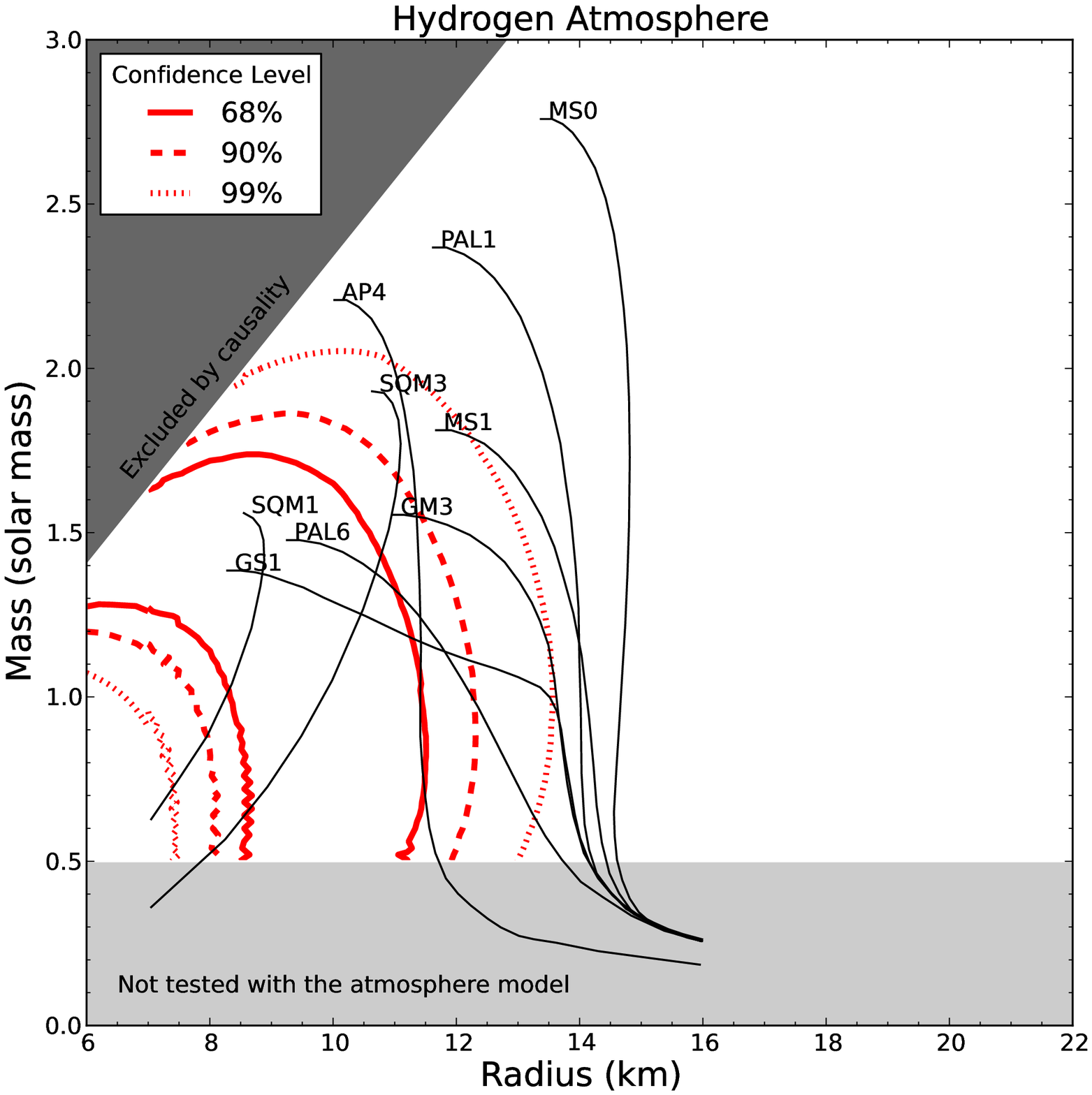}
\includegraphics[width=\columnwidth]{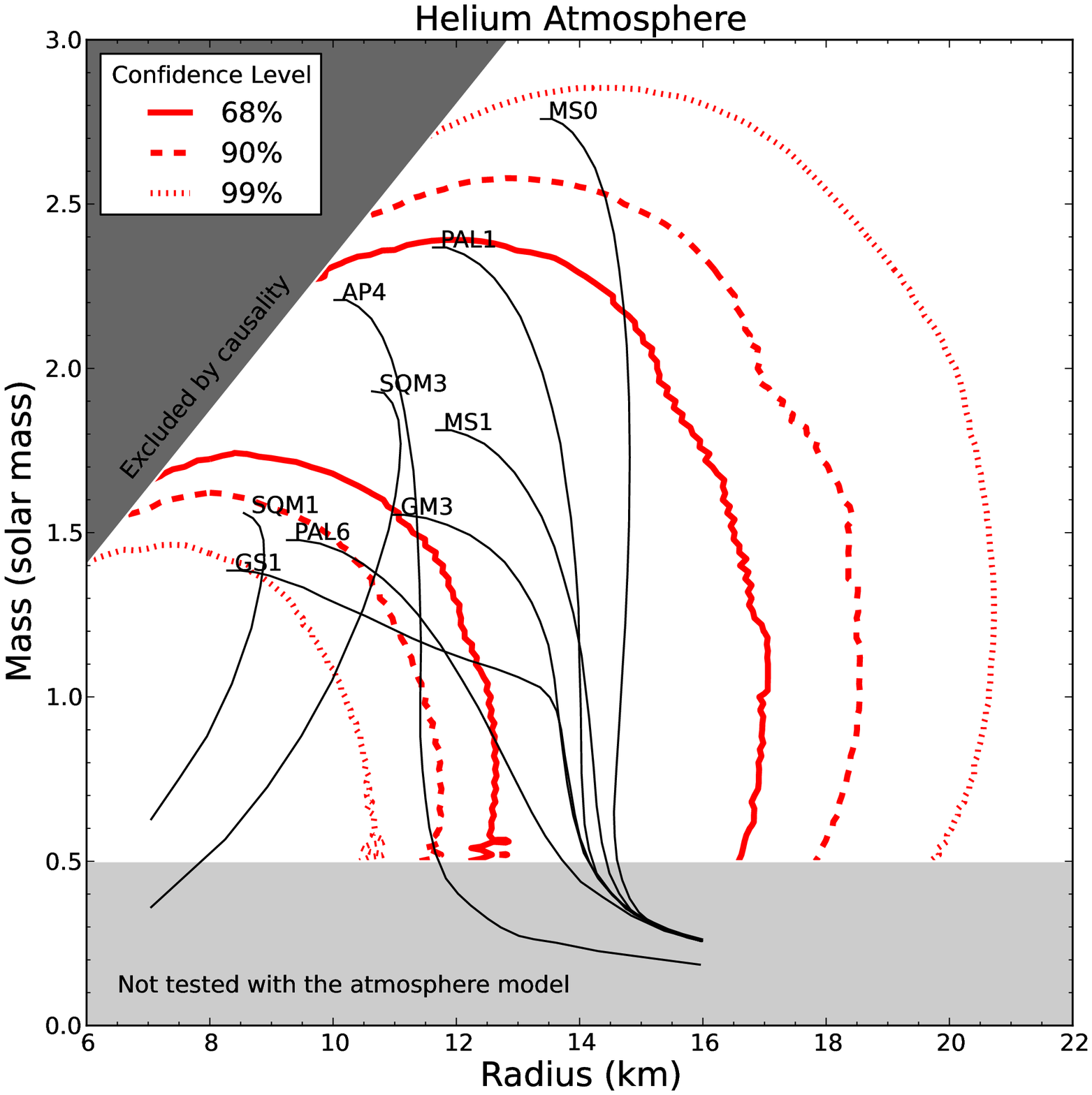}
\caption{Confidence levels for the mass and radius of the M28 NS \Chandra\ source 26, using a H (left) or He (right) atmosphere model. A representative selection of EOS are reported (labelled as in \citealt{Lattimer:2001p6716}). The parameters were not allowed to vary in the area ``Not tested with the atmosphere model''. We report in dark gray the area excluded by causality \citep{Rhoades:1974p7528}.}
\label{fig:contours}
\end{figure*}

Given the count rate and softness of the source, which is almost on-axis, we expect some level of pile-up in the spectrum (5\% using PIMMS 4.2\footnote{http://cxc.harvard.edu/toolkit/pimms.jsp}). This effect occurs when two or more photon events overlap in a single detector frame and are being read as a single event, creating a high energy tail in a contaminated spectrum. 
%We tried to add a power law component, which improved the fit ($\chi^{2}_{\nu}$/dof = 1.03/134) but did not flatten the residuals. 
We used the pile-up model component available in Xspec \citep{Davis:2001p1847} with a frame time set to 3.1~s and a free $\alpha$ parameter (related to the probability of events being retained as a good grade after filtering). This model gave the best fit ($\chi^{2}_{\nu}$/dof = 0.88/143), with ${\alpha=0.41\pm0.15}$.
% for an expected value of 0.5 to 0.7. 
%We froze $\alpha$ and 
We refitted the spectrum with the mass and radius parameters free to vary. The best fit model ($\chi^{2}_{\nu}$/dof = 0.87/141) is obtained for ${N_{\rm H}=(2.5\pm0.3)\times10^{21}}$~cm$^{−2}$, a temperature ${kT_{\rm eff}=125\pm40}$~eV, 
% 127 / 135 / 144
a mass ${M=1.4^{+0.4}_{-0.9}}$~\msun\ and a radius $R=9\pm3$~km. 
Errors are at 90\% significance {and we considered only masses higher than 0.5~\msun\ and radii higher than 6~km}. 
The 0.3--6~keV absorbed flux of the source (after removing the pile-up effect) is then ${(1.8\pm0.2)\times10^{-13}}$~\ergcm, and the unabsorbed luminosity ${\sim1.6\times10^{33}}$~\ergsec\ (at 5.5~kpc).
This spectrum is reported in Figure~\ref{fig:sp}.
We then ran the command \textit{steppar} and obtained confidence contours for the mass and radius of the NS, which are more instructive than the best fit parameter values and errors (see Figure~\ref{fig:contours}, left).

%\subsubsection{Spectral stability}

Though no obvious variability was found in section \ref{var}, we tested this assumption by untying the normalization for all five ObsIds, and only froze the ObsId 9132 value to 1. This led to the following best fit values for the normalization of each ObsId {with their 90\% confidence errors}: 
$0.93\pm0.08$ (2684), $0.89\pm0.08$ (2685), $0.99\pm0.08$ (2683), $1.0$
%\pm0.04$ 
(9132), $0.99\pm0.05$ (9133). 
{The largest deviation is thus at 2.3 sigma (ObsId 2685), still consistent with unity.}
We performed a similar test with $N_{\rm H}$ and found no significant variability as well. Consequently, we confirm that no significant variability is observed between the different epochs, both in the flux and in the spectral shape.

We also tested for the presence of a hard tail in the X-ray spectrum by adding a power law component. However, the normalization of this component quickly tends to zero, and the fraction of flux in the 0.3--8~keV energy band of this component cannot be higher than 5\% or 7\% of the total model flux (at the 95\% confidence level) for a photon-index of 1.5 and 2.5, respectively.

%\subsubsection{Helium and Carbon atmosphere models}

We performed the same fitting procedure 
%as in section~\ref{fitnsatmos} 
with an atmosphere model composed of pure He (using opacity tables computed by the Opacity Project\footnote{http://cdsweb.u-strasbg.fr/topbase/TheOP.html}; see \citealt{Ho:2009p7676} for details), and including the pile-up model.
A similar good fit was obtained ($\chi^{2}_{\nu}$/dof = 0.88/142, see Figure~\ref{fig:sp}) with ${N_{\rm H}=(2.65\pm0.25)\times10^{21}}$~cm$^{−2}$, a temperature ${kT_{\rm eff}=170^{+50}_{-90}}$~eV, 
% 80 / 172 / 219
a mass ${M=2.0^{+0.5}_{-1.5}}$~\msun\ and a radius $R=14^{+3}_{-8}$~km. 
The confidence contours obtained with the \textit{steppar} command are reported in Figure~\ref{fig:contours} (right). We note that the regions delimited by the contours are not consistent at the 80\% confidence level with the contours obtained with the H model.

Finally, we performed a similar fit with a carbon atmosphere model \citep{Ho:2009p7676}. We obtain an acceptable fit ($\chi^{2}_{\nu}$/dof = 0.88/142, see Figure~\ref{fig:sp}) but the parameter values are excluded by causality \citep{Rhoades:1974p7528}: $M>2.6$~\msun\ for $R=10\pm2$~km.

%--------------------
\section{Discussion}
\label{discuss}

%\subsection{Nature of the source}

%The \Chandra\ source 26 located in the GC M28 is a qLMXB candidate based on its luminous, soft X-ray emission well represented by a NS atmosphere model \citep{Becker:2003p734}.
The lack of variability of the source over six years agrees with the idea that the target is a LMXB in a quiescent, low-accretion state
%Indeed, long term variability over years is generally not observed for this kind of objects 
\citep[e.g.][]{Rutledge:2002p6660}. Such qLMXBs can be X-ray transients with episodes of higher accretion that would heat up the NS star surface to the current observed temperatures \citep{Brown:1998p7710}. Such a behavior has been observed for similar sources in GCs \citep[e.g.][]{Rutledge:2002p6656,Heinke:2003p7959,Heinke:2010p3183,Pooley:2006p1016,Degenaar:2011p7944}.
It is thus possible that this source becomes suddenly more luminous in the future.

%We obtained good fits with our NS atmosphere models for the X-ray spectra. The carbon atmosphere model can be rejected as the values of mass and radius obtained conflict with the causality principle. 
For both a H and a He model, we found good fits with an absorption consistent with the expected absorption from the GC reddening, suggesting that the source is located in the core of M28 with no or very low intrinsic absorption. The mass and radius are as expected for a typical NS \citep[e.g.][]{Lattimer:2010p8404}, and the temperature is in the expected range for qLMXBs.
No hard tail is observed in the spectrum, which seems to be a common feature of these objects in GCs (\citealt{Heinke:2003p7959}, but see \citealt{Wijnands:2005p7973}).

The only striking difference is that H and He atmosphere models give distinct contour regions of masses/radii at the 80\% confidence level (Figure~\ref{fig:contours}). On the one hand, the H model gives a mass and radius consistent with the canonical value of 1.4~\msun\ and 10~km, and allows for the presence of exotic matter inside NSs (hyperons, quarks). On the other hand, the He model provides solutions with higher masses/radii, consistent with the stiffest EOS for NS interiors, most of them composed of neutrons and protons.

%{One could argue that the He atmosphere model is preferred, given that the H model results favor EOS that do not reach a 2~\msun\ mass as observed for PSR J1614-2230 \citep{Demorest:2010p6698}.
% (except maybe EOSs similar to AP4).
%, which would give more credence to the He model.
{The accurate measurement of 1.97$\pm$0.04~\msun\ for PSR J1614-2230 \citep{Demorest:2010p6698} does not, by itself, indicate that a H atmosphere is ruled out since there are EOS that have a maximum mass above 2~\msun\ and lie within the H model confidence contours (AP4, but also others not shown in Figure~\ref{fig:contours}; see, e.g., \citealt{Lattimer:2010p8404,Weissenborn:2011p8410,Massot:2012p7975}).
The He atmosphere model also allows for several stiff EOS that reach masses higher than 2~\msun.}
%However, we report only a selection of EOS from \citet{Lattimer:2001p6716}, and more recent EOS overcome this apparent inconsistency \citep[e.g.][]{Weissenborn:2011p8410,Massot:2012p7975}.}
%Alternatively, EOSs similar to AP4 (see Figure~\ref{fig:contours}) are marginally consistent with the H atmosphere model results and could thus appear to be the only acceptable EOSs.}
%However, no firm conclusions can be drawn.
%given the loose constraints obtained on the mass and radius of the NS.
%{However, no firm conclusions can be drawn.} 
We note that there are additional sources of error to consider (beyond the statistical uncertainties from spectral fitting) that does not allow for strict constraints on the NS parameters and the EOS, and may slightly enlarge the regions showed in Figure~\ref{fig:contours}.
Uncertainties on the distance were not included in our analysis, but this value is relatively well constrained as the source is in a GC. This has no impact when comparing H and He models as the same distance is used for both models.
%{Part of the pile-up effect was accounted for in the fit, but it could induce additional uncertainties that are difficult to quantify.}
{Although pile-up is a small effect and was carefully modeled in the fit, the additional parameters of the pile-up model may introduce additional systematic uncertainties.}
%Also, the absorption on the line of sight is substantial and could affect the precision of the spectral modeling.

%\subsection{Atmosphere composition}

The composition of the NS atmosphere depends on the accreting material, physical processes occurring during the accretion, and conditions on the NS surface. 
%A non-evolved star would provide mainly H, while a white dwarf would provide heavier elements, mainly He. 
%Due to the strong surface gravity on the NS, only the lightest element remain in a pure, thin atmosphere.
%For NS in GCs, one might expect a higher rate of NS with a white dwarf companion, due to the dynamical evolution over a long timescale occurring in the core of GCs, and thus a higher chance of observing He NS atmospheres.
A non-evolved star will produce mostly H, which will quickly stratify to provide a pure H atmosphere.  White dwarf donors (in so-called ultra-compact LMXBs) will provide mostly He, C/O, or O/Ne/Mg depending on the white dwarf.  
Ultra-compact LMXBs are observed to be much more common in GCs than in the rest of the Galaxy \citep{Deutsch:2000p8022}.  Of 16 bright LMXBs in 13 clusters, we have 11 orbital period measurements, of which 5 indicate ultra-compact systems (e.g. \citealt{Zurek:2009p8046}; \citealt{Altamirano:2010p8084}). In the rest of the Galaxy, only 9 ultra-compact systems are known among the $\sim$80 bright LMXBs with period measurements (\citealt{Ritter:2003p2684,Ritter:2010p2694}).
%\footnote{http://physics.open.ac.uk/RKcat/}. 
Dynamical formation of ultra-compact LMXBs in GC cores explains this difference \citep{Verbunt:1987p8086,Ivanova:2005p8129}.

It is unclear whether spallation always produces H during accretion \citep{Bildsten:1992p7901,Bildsten:1993p7907,Chang:2004p8411}.
%(L. Bildsten 2008, priv. comm.; \citealt{Juett:2003p8131}).
Theoretical work is needed to clarify the conditions for spallation.  Obtaining a high-quality X-ray spectrum of an neutron star ultra-compact LMXB in quiescence at known distance would help clarify this question.  

%nature of the donor star for this (presumably transient) X-ray binary; if it's a He WD, and the accretion disk reaches the surface during outburst (so as to allow gentle slowing of the infalling matter and avoid spallation; if spallation occurs, then H may be produced in such a system, reference Bildsten et al. 1992), then a He atmosphere makes sense. \cite{Bildsten:1992p7901}

%A He atmosphere model could better explain the emission of GC qLMXBs reported with low masses.
Among the qLMXBs in GCs that were used to derive constraints on the mass and radius of their NS using a H atmosphere models, some were reported to have a low mass or radius. 
For example, the source in the core of NGC 6553 (XMMU J180916−255425) gives a NS radius $R_{\rm NS}=6.3^{+2.3}_{-0.8}$~km for a mass of 1.4~\msun\ \citep{Guillot:2011p6830}. 
The qLMXB in M13 \citep{Gendre:2003p1003} and the one in NGC 2808 \citep{Servillat:2008p2227,Servillat:2008p2247} also showed relatively low masses/radii with a radius lower than 10~km or a mass around 1~\msun\  \citep{Webb:2007p649,Steiner:2010p7695}, however the values are not well constrained.
In a same way, the qLMXB U24 in NGC 6397 was reported to show a relatively small radius $R_{\rm NS} = 8.9^{+0.9}_{-0.6}$~km for a mass of 1.4~\msun\ \citep{Guillot:2011p6829}.
On the other hand, X7 in 47 Tuc was reported with higher values of mass/radius \citep[$R>12$~km or $M>2$~\msun,][]{Heinke:2006p1069}, while the NS in $\omega$~Cen appears as an intermediate case \citep{Rutledge:2002p6656,Gendre:2003p483}, as is the NS studied in this work.

Following our study of the qLMXB in M28, it is possible that some of those sources harbor a NS with a He atmosphere, rather than a H atmosphere. This would favor higher radii and masses for NS, and thus stiffer EOS, in agreement with the precise measurement of relatively high masses for some NS \citep[e.g. $\sim$2~\msun,][]{Demorest:2010p6698}.
We will thus try fitting other quiescent LMXBs with He (and carbon) atmospheres in future work.

Identifying the composition of the atmosphere of known quiescent LMXBs is clearly of key importance, and we suggest three means of doing so. 
i) Spectroscopy, or (less time-consuming) narrow-filter photometry of optical counterparts can identify H$\alpha$ emission from LMXBs in quiescence or outburst and thus the presence of H; the LMXB in $\omega$~Cen \citep{Haggard:2004p8145} and X4 and X5 in 47~Tuc (van den Berg et al., in prep) therefore possess H atmospheres.  
ii) Orbital periods differentiate between ultra-compact and longer period systems; we note that long periods are known for X5 and W37 in 47 Tuc \citep{Heinke:2005p8154}, suggesting a main-sequence companion and accretion of H.
iii) Finally, thermonuclear bursts can distinguish between H-rich and H-poor environments, particularly at low (${<0.01~\dot{M}_{\rm Edd}}$) accretion rates where H should burn unstably (e.g. \citealt{Fujimoto:1981p8215}; \citealt{Galloway:2008p8258}).  

This last point is of particular interest for the M28 qLMXB, since a peculiar X-ray burst was observed from this GC \citep{Gotthelf:1997p8264}.  This burst was unusually low-luminosity, suggesting burning on only one patch of the star. The short timescale of this burst ($\tau$=7.5~s) requires He burning without the presence of H, and thus (given the quiescent state) pure He accretion and a pure He atmosphere. Unfortunately we cannot be certain that this burst originated from the known qLMXB, as other qLMXBs may be hidden among the fainter sources in this cluster.

%--------------------
%\section{Conclusion}

\section*{Acknowledgments}

We thank the referee for their careful reading and useful suggestions that strengthened the paper. MS acknowledges supports from NASA/Chandra grant GO0-11063X, NSF grant AST-0909073 and the Centre National d'Etudes Spatiales (CNES). COH is supported by NSERC and an Ingenuity New Faculty Award. WCGH appreciates the use of the computer facilities at KIPAC and acknowledges support from STFC in the UK.

\bibliographystyle{mn2e} 
\bibliography{../../ref}

\label{lastpage}

\end{document}